\documentclass{elsart1}
\usepackage{epsfig}

\begin{document}
\begin{frontmatter}
\title{\large\bf\boldmath Search for the Rare Decays $J/\psi \rightarrow D_{S}^{-} \pi^{+}$,
$J/\psi \rightarrow D^{-} \pi^{+}$ and $J/\psi \rightarrow \bar
D^{0} \bar K^{0}$}

\begin{small}
\begin{center}
\vspace{0.2cm}

M.~Ablikim$^{1}$,              J.~Z.~Bai$^{1}$, Y.~Ban$^{12}$,
X.~Cai$^{1}$,                  H.~F.~Chen$^{17}$,
H.~S.~Chen$^{1}$,              H.~X.~Chen$^{1}$, J.~C.~Chen$^{1}$,
Jin~Chen$^{1}$,                Y.~B.~Chen$^{1}$
% ...
\\
\vspace{0.2cm}
(BES Collaboration)\\
\vspace{0.2cm}
{\it

$^{1}$ Institute of High Energy Physics, Beijing 100049, People's Republic of China\\
% ...
$^{12}$ Peking University, Beijing 100871, People's Republic of China\\
% ...
$^{17}$ University of Science and Technology of China, Hefei 230026, People's Republic of China\\
% ...
}

\end{center}

\end{small}
\maketitle
\normalsize
\begin{abstract}

~~Rare decay modes $J/\psi \rightarrow D_{S}^{-}
\pi^{+} + c.c.$, $J/\psi \rightarrow D^{-} \pi^{+} + c.c.$ and $J/\psi \rightarrow
\bar D^{0} \bar K^{0} + c.c.$ are searched on a basis of 
5.77$\times 10^{7}$ $J/\psi$ collected with the BESII detector at BEPC. 
No signal
above background is observed. We present upper
limits on the branching fraction
$B(J/\psi \rightarrow D_{S}^{-} \pi^{+})$ $<$
1.4$\times$10$^{-4}$,
$B(J/\psi \rightarrow D^{-} \pi^{+})$ $<$7.5$\times$10$^{-5}$
and $B(J/\psi \rightarrow \bar
D^{0} \bar K^{0})$ $<$ 1.7$\times$10$^{-4}$ with 90$\%$ confidence level.

\end{abstract}
\end{frontmatter}

\section{INTRODUCTION}
\label{secintro}

\begin{figure}[htbp]
  \centering
  \begin{minipage}[b]{0.4\linewidth}
  \centering
  \includegraphics[height=3.0cm,width=4.5cm]{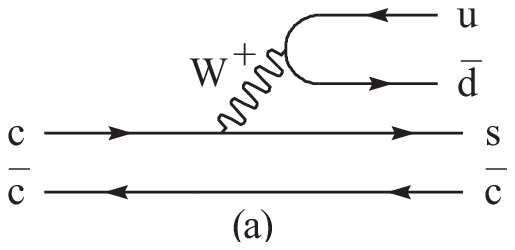}
  \end{minipage}
  \begin{minipage}[b]{0.4\linewidth}
  \centering
  \includegraphics[height=3.0cm,width=4.5cm]{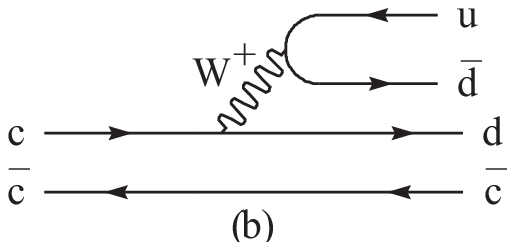}
  \end{minipage}
  \begin{minipage}[b]{0.4\linewidth}
  \centering
  \includegraphics[height=3.0cm,width=4.5cm]{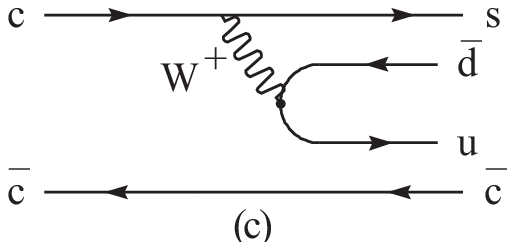}
  \end{minipage}
  \caption{
  Leading order electroweak Feynman diagrams for (a) $J/\psi \rightarrow D^-_S
    \pi^+$, (b) $J/\psi \rightarrow
      D^- \pi^+ $, and (c) $J/\psi \rightarrow \bar D^0 \bar K^{0}$ in the standard model.}
\label{feynme}
\end{figure}

~~The hadronic decays and electromagnetic decays of $J/\psi$ have
been widely studied, however weak decays of $J/\psi$ 
have not been studied in detail~\cite{sunss}.  
For the $J/\psi$ lying below the $D \bar D$
threshold, decays to $D \bar D$ are forbidden. Weak decays to one
single charm meson accompanied by other non-charm mesons are allowed.  
The standard model
predicts that the branching fractions of flavor changing processes
via weak interactions, such as $J/\psi \rightarrow D X$ (where $D$ stands for
$D_S$ or $D$ and $X$ stands for $\pi$ or $K^{0}_{S}$) are at the level 
of 10$^{-8}$ or below~\cite{zphy} and thus
smaller than those of strong and electromagnetic decays. 
This is unobservable in
current experiments. Extensions to the standard model, such as Top Color
models~\cite{topcol}, the minimal super-symmetric standard model
with or without R-parity~\cite{susy} and the two Higgs doublet
model~\cite{2hdm}, can enhance the branching fraction to
$\sim$10$^{-5}$. The study of $J/\psi \rightarrow D_{S}^{-}
\pi^{+}$, $J/\psi \rightarrow D^{-} \pi^{+}$ and $J/\psi \rightarrow
\bar D^{0} \bar K^{0}$ provide an experimental 
check and serve as a probe to new physics~\cite{zhangxm,datta}.  Figure
\ref{feynme} shows the Feynman diagrams for these decay modes in
the framework of the standard model, respectively. 
Charge
conjugated states are implicitly included.
In this paper, we'll
present a search for $J/\psi \rightarrow D_{S}^{-} \pi^{+}$, $J/\psi
\rightarrow D^{-} \pi^{+}$ and $J/\psi \rightarrow \bar D_{0} \bar
K^{0}$ in a sample of 5.77$\times 10^{7}$ $J/\psi$ events collected
with the Beijing Spectrometer(BESII)~\cite{besii} detector at the Beijing
Electron-Positron Collider(BEPC)~\cite{bepc}. 

\section{BESII Detector}
\label{besdet}

~~BES is a conventional solenoidal magnetic detector that is
described in detail in Ref.~\cite{bes}. BESII is the upgraded
version of the BES detector ~\cite{besii}. A 12-layer Vertex Chamber
(VC) surrounding the beryllium beam pipe provides track and
trigger information. A forty-layer main drift chamber (MDC) located
just outside the VC provides measurements of charged particle
trajectories covering $85\%$ of 4$\pi$; it also provides
ionization energy loss ($dE/dx$) measurements which are used for
particle identification (PID).  A momentum resolution of
$0.017\sqrt{1+p^2}$ ($p$ in GeV/$c$) and a $dE/dx$ resolution for
hadronic tracks of $\sim8\%$ are obtained.  
Time of flight (TOF) of charged particles is measured with an array of 48 scintillation
 counters surrounding the MDC. 
The resolution is about 200 ps
for hadrons. Outside the TOF counters, a 12 radiation length,
lead-gas barrel shower counter (BSC),
 measures energies and positions of electrons and
 photons. Solid angle is covered over $80\%$ 
 and resolutions of
 $\sigma_{E}/E=0.22/\sqrt{E}$ ($E$ in GeV), $\sigma_{\phi}=7.9$
 mrad, and $\sigma_{z}=2.3$ cm are obtained. Outside the solenoidal coil,
 providing a 0.4 T magnetic field over the tracking volume, an
 instrumented flux return with three double-layer muon
 counters identify muons with momenta greater than 500~
 MeV$/c$.

 Monte Carlo simulations are performed using a GEANT3 based program (SIMBES) with
 detailed consideration of the detector geometry and response. The
 consistency between data and Monte Carlo has been checked in many
 reactions in $J/\psi$ and $\psi(2S)$ decays with
 reasonable agreement. Details are described in
 Ref.~\cite{simbes}.\par

\section{Event Selection}
\label{gsecevt}

~~Due to a large $J/\psi$ hadronic decay background,  non-leptonic decay
modes do not offer ultimate sensitivity. In this analysis, $D_{S}$ and
$D$ mesons are reconstructed via
semileptonic decay modes: $D_{S}^{-} \rightarrow \phi e^{-} \nu_{e}$,
$D^{-} \rightarrow K^{*0} e^{-} \nu_{e}$ and $D^{-} \rightarrow
K^{0} e^{-} \nu_{e}$, $\bar D^{0} \rightarrow K^{+} e^{-} \nu_{e}$.
The neutrino is undetectable in the detector, but carries energy
and momentum. $D_{S}$ and $D$ mesons can not be identified by their
invariant mass. However, two body constraints can be applied in the mode 
$J/\psi
\rightarrow D X$. Thus they are identified using the momentum
information of $X$ meson.

~~Four charged tracks are required in all selected modes, and the
total charge must be equal to zero. In order to ensure
well-measured momenta and reliable particle identification, all
the tracks are required to be reconstructed in the main drift
chamber with a good helix fit. Each track is required to
 satisfy the geometry selection criterion $|\cos\theta|<0.8$
, where $\theta$ is the polar angle, and must originate from the beam
interaction region (except the decay daughters of $K^{0}_{S}$), which is defined by $R_{xy}
< 2.0 $cm and $|z| < 20.0 $cm, where $R_{xy}$ and $|z|$ are the
closest approach of the charged track in the xy plane and z
direction.

~~A kaon or pion candidate is required to satisfy $W_{K,\pi}>0.1\%$, where $W_{K,\pi}$ 
is the weighted likelihood of the hypothesis of kaon or pion, which combines the TOF and 
$dE/dx$ information. To reduce the misidentification rate, the likelihood ratio 
$R_{K,\pi} = W_{K,\pi}/(W_{K}+W_{\pi})$ for kaon or pion is required to be greater than 
0.7. The ratio of the energy deposit in the BSC to the momentum is 
used to construct the likelihood for electron identification, $W_{e} > 1\%$ and 
$R_{e}=W_{e}/(W_{\pi}+W_{K}+W_{\pi}) > 0.85$ are required for electron candidate.
The angle
between the identified electron and the nearest charged track is
required to be greater than 12$^{\circ}$ to remove backgrounds from
photon conversion. Low momentum electrons and pions can
not be unambiguously identified, leading to a momentum cut
 $P_{e}>0.25$~GeV/c for electrons.

 ~~Neutral Kaons are identified via the decay to $\pi^{+}$$\pi^{-}$.
 All pairs of oppositely charged tracks are assumed to be
$\pi^{+}$$\pi^{-}$. The distance between the $K^{0}_{S}$ decay
vertex and the beam axis is required to exceed 5 mm the in xy
plane. Background from $J/\psi$'s decaying to
states with extra neutral particles is removed by constraining the number of 
isolated photons to be zero. An isolated photon is a photon with an
angle between the nearest charged track and the cluster of at least
22$^{\circ}$ and a difference between the angle of the cluster
development direction in the BSC and the reconstructed photon emission direction
of less than 60$^{\circ}$. In addition the energy deposit in the BSC
is required to be larger than 0.1~GeV.

~~The neutrino of the semi-leptonic decay of the reaction $J/\psi\rightarrow DX$
remains undetected.
A kinematic quantity $U_{miss} = E_{miss} - P_{miss}$ is
used to identify missing neutrinos,
 where $E_{miss}$ and $P_{miss}$ are the energy
and momentum of the neutrino. Ideally $U_{miss}$
should be consistent with zero. A requirement of $|U_{miss}| < 0.1$
~GeV  may remove backgrounds from $J/\psi$ decaying to $K^{0}_{L}$,
$\eta$ and partial $\pi^{0}$ final states which can not be rejected
through the initial selection. 
Events with missing energy due to misidentified pions are rejected by 
$P_{miss}>0.2$~GeV/c. Missing particles are required to be in the sensitive region 
  of the detector to further supress hadronic background.

\begin{figure}[htbp] \centering
\begin{minipage}[b]{0.4\linewidth}
\centering
\includegraphics[height=6.0cm,width=5.5cm]{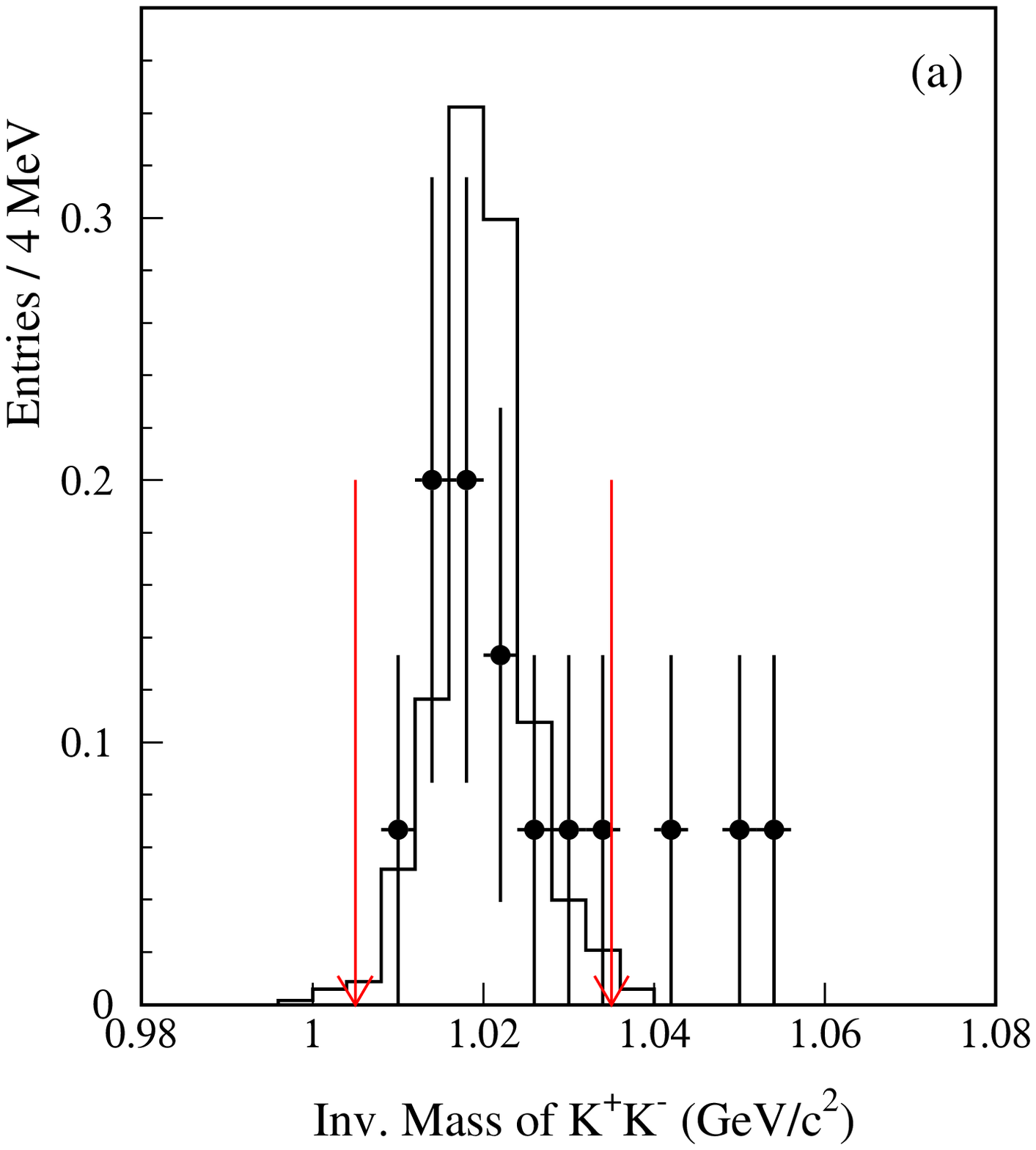}
\end{minipage}
\centering
\begin{minipage}[b]{0.4\linewidth}
\includegraphics[height=6.0cm,width=5.5cm]{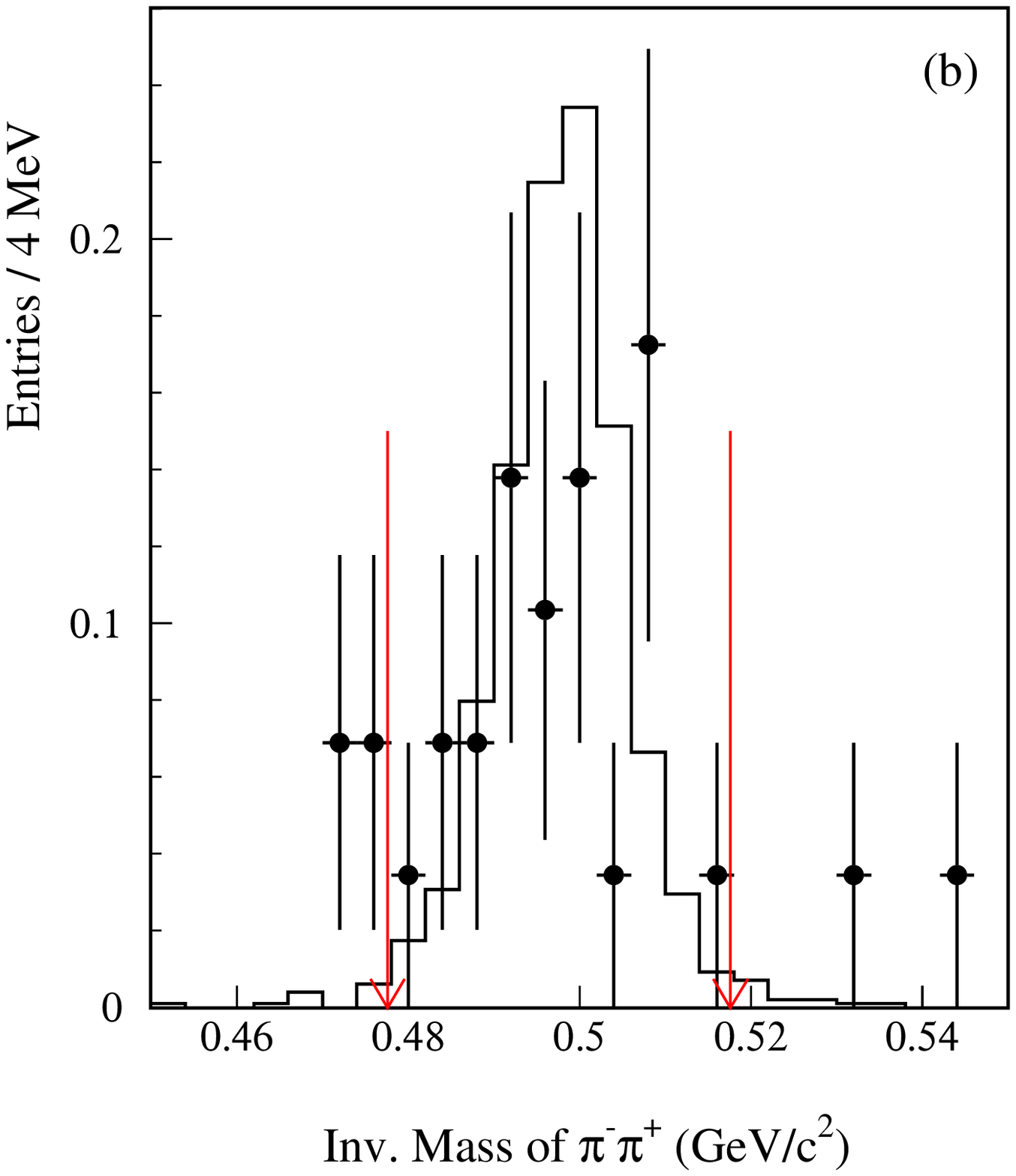}
\end{minipage}
\begin{minipage}[b]{0.4\linewidth}
\centering
\includegraphics[height=6.0cm,width=5.5cm]{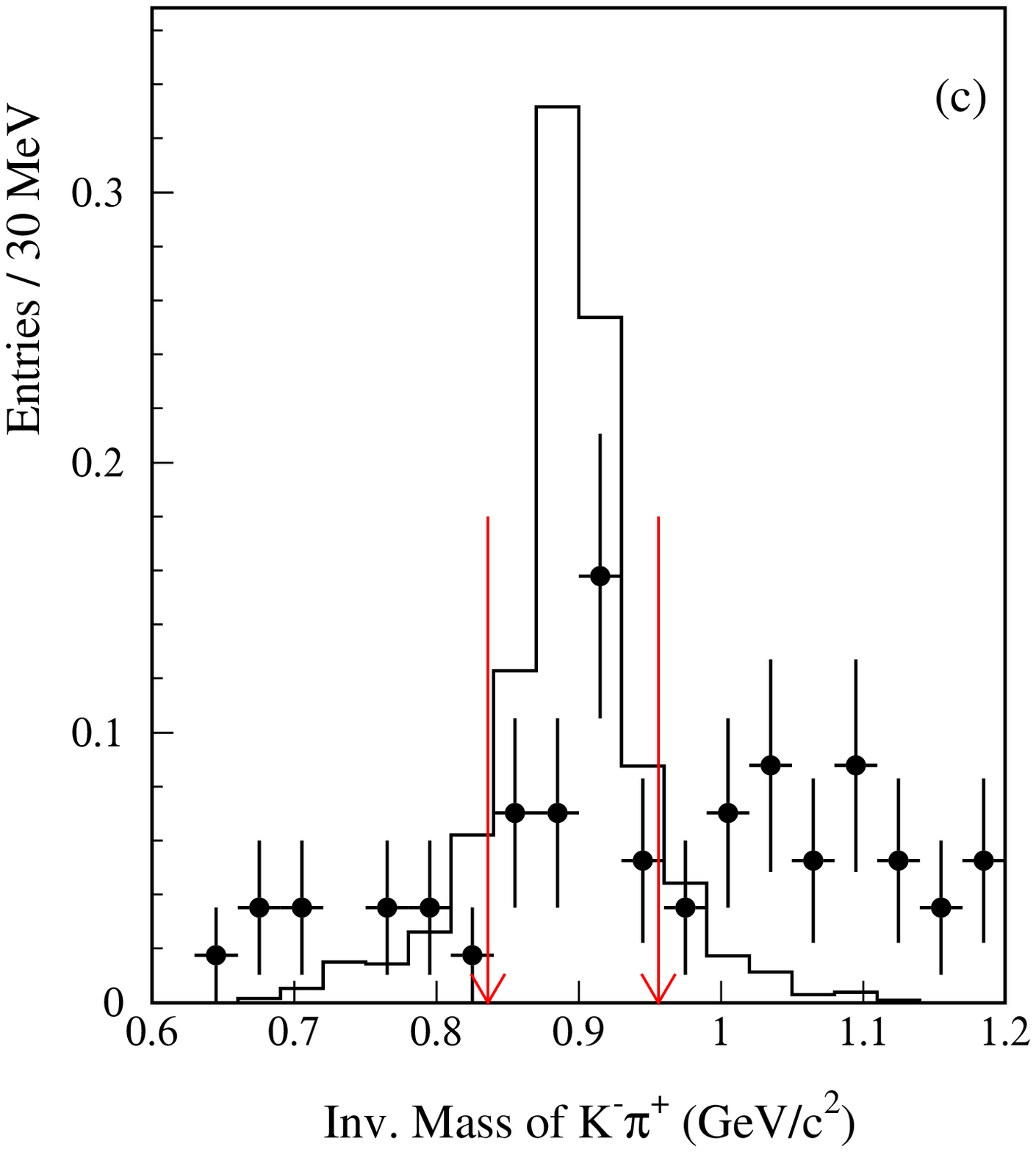}
\end{minipage}
\begin{minipage}[b]{0.4\linewidth}
\centering
\includegraphics[height=6.0cm,width=5.5cm]{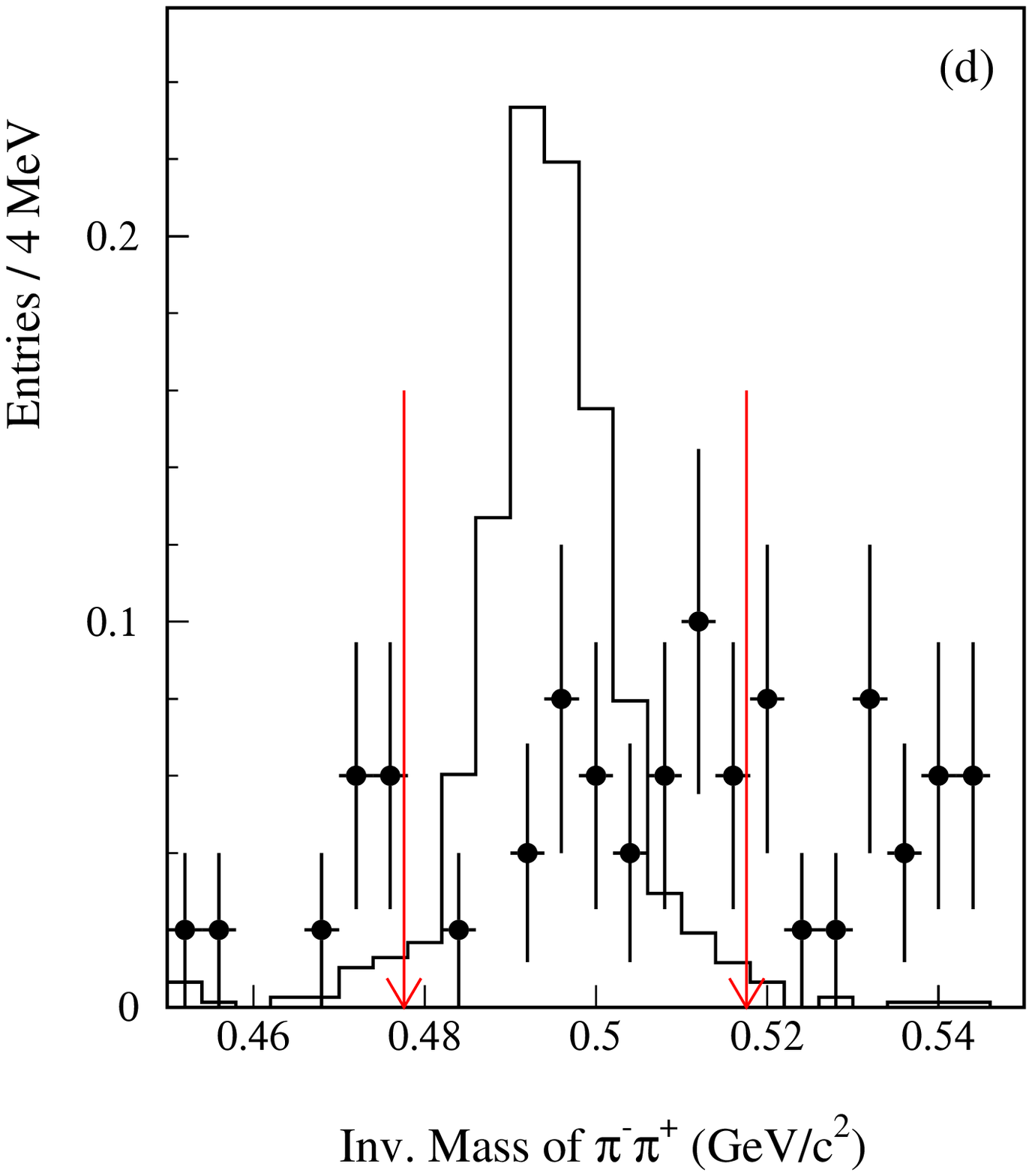}
\end{minipage}
\caption{The invariant mass distribution of resonance
candidates  (a) $\phi$ from $J/\psi \rightarrow D^-_S \pi^+
$, (b) $K^{0}_{S}$ from $J/\psi \rightarrow \bar {D^0} \bar K^{0}$,
(c) $K^{*0}$ from $J/\psi \rightarrow
D^- \pi^+$, $D^- \rightarrow K^{0*} e^{-} {\nu}_e$, and (d)$K^{0}_{S}$ from
$J/\psi
\rightarrow D^{-} \pi^+$, $D^- \rightarrow K^{0} e^- {\nu}_e$. 
Data is shown as dots with error bar,
the expected signal shape from Monte Carlo simulated signal events is shown as a histogram. 
The mass cuts are illustrated by arrows and are discussed in the text. 
}
\label{res}
\end{figure}

~~In the decay mode $J/\psi \rightarrow D^{-}_S \pi^+$, $D^{-}_{S}$ mesons are
reconstructed through their decay to $D^{-}_{S} \rightarrow \phi e^{-} \nu_{e}$,
$\phi$ candidates are
reconstructed from two oppositely charged kaons, namely:
 $\phi \rightarrow K^{+}K^{-}$.
 The invariant mass of $\phi$ candidates is required to be within 0.015 GeV/$c^2$ of
 the $\phi$ nominal mass.
  ~~In the decay mode $J/\psi \rightarrow D^-\pi^+$,
   $D^{-}$ mesons are reconstructed through their decays to
 $D^{-} \rightarrow  K^{*0} e^{-} \nu_{e}$
      and $D^{-} \rightarrow  K^{0} e^{-} \nu_{e}$.
      The $ K^{*0}$ candidates are formed out of $K^-$ and ${\pi}^{+}$ candidates.
       Their invariant mass is required to fulfil
       $|M_{K\pi}-M_{K^{*0}}|<0.060$~GeV/$c^2$.
  The $K^{0}_{S}$ candidates are formed out of $\pi^{+}$ and $\pi^{-}$ candidates. The mass
    window is $|M_{\pi^{+}\pi^{-}}-M_{K^{0}_{S}}|<0.020$~GeV/$c^2$.
      As far as the decay mode $J/\psi \rightarrow \bar D^{0} \bar K^{0}$ is
concerned, $\bar D^{0}$ mesons are reconstructed through their decay to $ K^{+} e^{-} \nu_{e}$. 
 ~~Figure \ref{res} shows
the invariant mass distribution of
$KK$, $ K\pi$ and $\pi\pi$ from the various decay modes.

\begin{figure}[htbp]
\centering
\includegraphics[height=9.0cm,width=7.5cm]{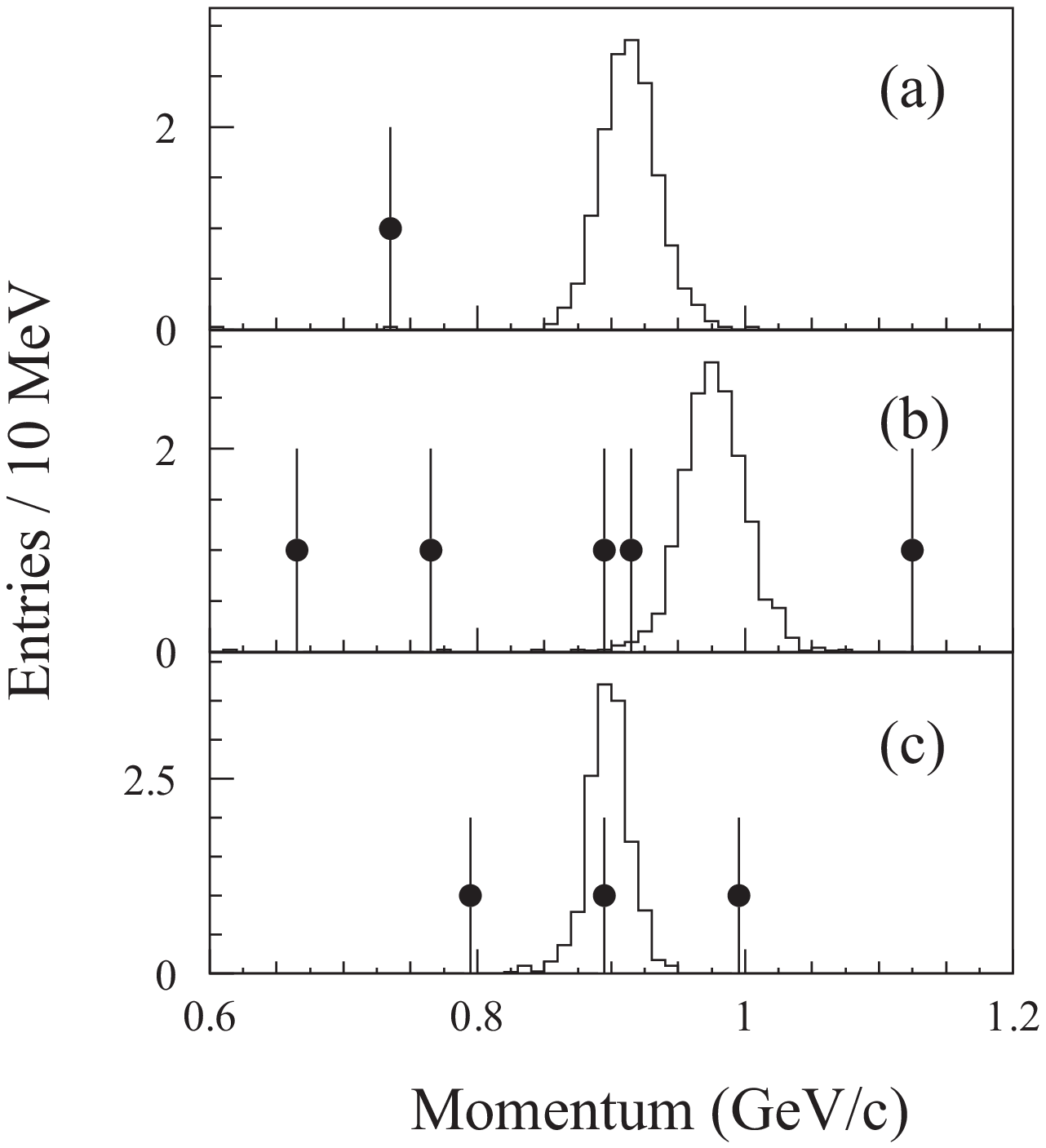}
\caption{
Momentum distributions of recoil $D_{S}$ or $D$ meson. (a) $J/\psi \rightarrow
D^-_S \pi^+
$, (b) $J/\psi \rightarrow D^- \pi^+$,
(c) $J/\psi \rightarrow \bar {D^0} \bar K^{0}$.
Data is shown as dots with error bars. The expected signal shape from 
Monte Carlo simulated signal events is shown as a histogram.}
\label{result}
\end{figure}

~~Figure \ref{result} shows the recoil momentum distribution of
$D_{S}$ or $D$ meson for the three decay modes after all selection criteria.
The two decay modes of $D^- \rightarrow K^{0*} e^{-} {\nu}_e$ and
 $D^- \rightarrow K^{0} e^- {\nu}_e$ are combined in Fig.\ref{result}(b).

\vspace{0.5cm}

\section{Monte Carlo Simulation}
\label{mcsim}

~~Detection
efficiencies are determined by detailed Monte Carlo simulations. 
$J/\psi \rightarrow D^{-}_S
\pi^+$, $J/\psi \rightarrow D^-\pi^+$ and $J/\psi \rightarrow \bar
D^{0} \bar K^{0}$ production and decay, including the detector response are simulated. 
50,000 Monte Carlo events are generated for each decay mode.
Branching fractions of $D^-_S$, $D^-$,
and $\bar {D^0}$ decays are taken from the world average values~\cite{PDG}.
 The detection
efficiencies and branching fractions are listed in Table 1.

\begin{table}[htbp]
  \centering
 \caption{Detection efficiencies and branching fractions~\cite{PDG}.}
 \begin{tabular}{|c|c|c|c|} \hline   \hline
     Decay mode & Intermediate decay&${\varepsilon}$  & ${B}$  \\
                \hline
 $J/\psi\rightarrow D^-_S \pi^+$  &$D^-_S\rightarrow \phi e^- \nu_{e}, \phi \rightarrow K^{-}K^{+}$& 2.46\% & $(2.4{\pm}0.4)\% \times
      (49.2{\pm}0.6)\%$ \\
          \hline
 $J/\psi\rightarrow D^- \pi^+$ & $D^-\rightarrow K^{*0} e^- \nu_{e}, K^{*0} \rightarrow K^{-}\pi^{+}$ &  1.90\% & $(3.74{\pm}0.21)\%$
      \\   \hline
     $J/\psi\rightarrow D^- \pi^+$& $D^-\rightarrow K^{0} e^- \nu_{e},  K^{0}_{S} \rightarrow \pi^{+}\pi^{-}$ &  3.76\% & $(8.6{\pm}0.5)\% \times 1/2
      \times (69.20{\pm}0.05)\%$ \\
           \hline

  $J/\psi\rightarrow \bar D^{0}\bar K^{0}$&$\bar {D^0}\rightarrow K^+ e^- \nu_{e}
  , K^{0}_{S} \rightarrow \pi^{+}\pi^{-}$ &  4.82\% & $(3.51{\pm}0.11)\% \times 1/2
  \times (69.20{\pm}0.05)\%$  \\
                                        \cline{2-3}
    \hline
 \end{tabular}
 \end{table}

\section{Systematic Errors}
\label{syserr}

~~The systematic error of the branching fraction
is dominated by uncertainties of MDC
simulation (including systematic uncertainties of the tracking efficiency and 
other requirements).
The efficiency varies between 14.2\% and 19.9\%.
We estimate a systematic error of $2\%$ for requiring the number of
isolated photons to be equal to zero,
 using $J/\psi\rightarrow{\rho}{\pi}$ decays~\cite{gam}.
The systematic uncertainty of false 
 electron identification is estimated to be $5\%$ for each
 electron. 
 Differences in the pion and kaon identification between Data and Monte Carlo 
  simulation indicate a systematic error of 1.5$\%$ for each track.
The errors on the
 intermediate decay branching fractions of
$D^-_S$, $D^-$, $\bar {D^0}$, $\phi$,
and $K^{*0}$ are taken from the PDG~\cite{PDG}. The statistical
error of the Monte Carlo sample is also taken into account. The
total number of $J/\psi$ events is
$(57.7{\pm}2.7){\times}10^{6}$~\cite{jpsinum}, determined from
inclusive 4-prong hadronic final states, and 4.7\%
is taken as a systematic uncertainty. All systematic errors 
are summarized in Table 2.

\begin{table}[htbp]
\scriptsize
\centering
\caption{Summary of the systematic errors.}
\begin{tabular}{|c|c|c|c|c|}
    \hline
    \hline
   & {\scriptsize $J/\psi\rightarrow{D}^{-}_{S}\pi^{+} $} &\multicolumn{2}{c|}
       {\scriptsize $J/\psi \rightarrow D^-
  \pi^+ $} & {\scriptsize $J/\psi
  \rightarrow \bar {D^0} \bar K^{0}$} \\\hline
& \scriptsize ${D}^{-}_{S}\rightarrow\phi e^- {\nu}_{e}$ & \scriptsize ${D}^{-}
  \rightarrow{K}^{*0}e^-{\nu}_{e}$ &\scriptsize ${D}^{-} \rightarrow{K}^{0}e^-{\nu}_{e}$
  & \scriptsize $\bar{D^0}\rightarrow{K}^+ e^- {\nu}_{e}$
 \\
 \hline
 \scriptsize MDC Simulation & {15.6\%} & 14.2\% &
       {19.9\%} &15.5\% \\
                                     \hline
      \scriptsize Photon veto    & 2.0\% & 2.0\%  & 2.0\% & 2.0\% \\
     \hline
   \scriptsize $e$ PID    & 5.0\% & 5.0\% & 5.0\%& 5.0\% \\
     \hline
   \scriptsize $\pi$, $K$ PID & 4.5\% & 4.5\% & 4.5\% & 4.5\% \\
    \hline
   \scriptsize $B(D_s,D)$ & 16.7\% & 5.6\% & 5.8\% & 3.1\%  \\
    \hline
   \scriptsize MC Statistics & {4.0\%} & 3.2\% & {3.9\%}&3.4\% \\
    \hline
    \scriptsize number of $J/\psi$    & 4.7\% & 4.7\% & 4.7\%& 4.7\% \\
 
    \hline
   \scriptsize total    & 24.7\% & 17.7\% & 22.7\% & 18.3\%  \\
    \hline
   \hline
  \end{tabular}
  \end{table}

\section{Results}
\begin{figure}[htbp]
\centering
\includegraphics[width=7.0cm]{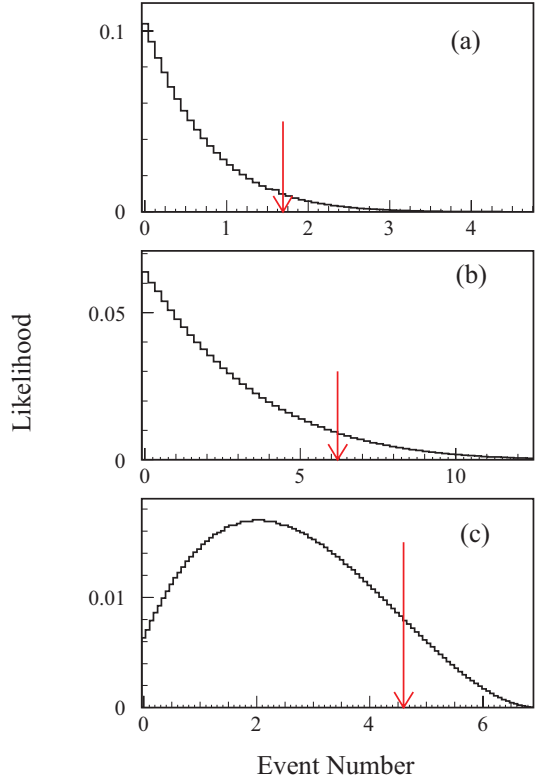}
\caption{Likelihood distributions for the observed number of
        events for (a) $J/\psi\rightarrow D^-_{S} \pi^+ $,
    (b) $J/\psi\rightarrow D^- \pi^+ $, and (c) $J/\psi \rightarrow
        \bar {D^0} \bar K^{0}$. The observed number of events at
          a Bayesian 90\% confidence level for the three channels are
            indicated by arrows in the plots.}
\label{uplim}
\end{figure}

~~Since no excess is observed for the reactions $J/\psi\rightarrow{D}^{-}_{S} \pi^+ $,
 $J/\psi \rightarrow D^{-} \pi^{+} $ and $J/\psi\rightarrow
 \bar{D^0} \bar K^{0} $ is observed above background, upper
 limits for the three decays are calculated. 
Using a Bayesian method, the upper limits for the observed number of events
 at the 90\%
 confidence level are  1.70 for $J/\psi\rightarrow{D}^{-}_{S} \pi^+ $, 6.21
 for $J/\psi \rightarrow D^{-} \pi^{+} $ and 4.61 for
 $J/\psi\rightarrow \bar{D^0} \bar K^{0} $ as shown in Fig.\ref{uplim}.

~~The upper limits on branching fractions are calculated by
\begin{equation}
B_{up}<\frac{n^{obs}_{UL}}{N_{J/\psi}{\varepsilon}B(1-{\sigma}^{sys})}
\end{equation}
where $n^{obs}_{UL}$ is the upper limit of the observed number of
events at the 90\% confidence level. $N_{J/\psi}$ is the total number of
$J/\psi$ events,  ${\varepsilon}$ and $B$ are the detection efficiency
and branching fraction, respectively. For the decay mode  $J/\psi \rightarrow
D^- \pi^+$, ${\varepsilon}B$ is the sum of the products obtained from decay
modes
$D^- \rightarrow K^{0*} e^{-} {\nu}_e$ and
 $D^- \rightarrow K^{0} e^- {\nu}_e$.
${\sigma}^{sys}$ is the systematic error of the three decay modes,
in the range of 18.3\% to 24.7\%. In the decay mode $J/\psi\rightarrow{D}^{-}\pi^+$, 
$D^{-}$ mesons are reconstructed from two decay modes,
the systematic uncertainty is 21.2$\%$ weighted by their branching fractions.

\begin{table}[htbp]
\centering \caption{Numbers used in the calculation of upper limits
on the branching fractions of $J/\psi \rightarrow D^{-}_S
 \pi^{+}$, $J/\psi \rightarrow D^{-} \pi^{+}$
   and $J/\psi\rightarrow \bar D^0 \bar K^{0} $.} {\small
\begin{tabular}{|c|c|c|c|}
       \hline
       \hline
       & $J/\psi\rightarrow{D}^{-}_{S}\pi^+ $ & $J/\psi
         \rightarrow D^-
    \pi^+ $ & $J/\psi \rightarrow \bar D^0 \bar K^{0} $\\
       \hline
         $n^{obs}_{UL}$ & 1.70 & 6.21 & 4.61  \\
       \hline
        ${\varepsilon}{B}$ & $2.90{\times}10^{-4}$ & $1.83{\times}10^{-3}$ & $5.85{\times}10^{-4}$ \\
        \hline
            Sys. Err. & 24.7\% & 21.2\% & 18.3\% \\
        \hline
          $B$(90\%C.L.) & $<1.4{\times}10^{-4}$ &
        $<7.5{\times}10^{-5}$ & $<1.7{\times}10^{-4}$ \\
        \hline
        \hline
 \end{tabular}
 }
\end{table}

In summary, we have searched for the decays of $J/\psi \rightarrow
D^-_S \pi^+ $, $J/\psi \rightarrow D^- \pi^+ $ and
$J/\psi \rightarrow \bar {D^0} \bar K^{0} $ using $5.77 \times 10^7
J/\psi$ events taken by the BESII detector at the BEPC $e^+
e^-$ collider. No evidence for any of these decays is found. The
final results for the 90\% confidence level upper limit of the
branching fractions are listed in Table 3.

{Acknowledgements}
\vspace{0.4cm}\\

The BES collaboration thanks the staff of BEPC and computing
center for their hard efforts. This work is supported in part by
the National Natural Science Foundation of China under contracts
Nos. 10491300, 10225524, 10225525, 10425523, 10625524, 10521003,
the Chinese Academy of Sciences under contract No. KJ 95T-03, the
100 Talents Program of CAS under Contract Nos. U-11, U-24, U-25,
and the Knowledge Innovation Project of CAS under Contract Nos.
U-602, U-34 (IHEP), the National Natural Science Foundation of
China under Contract No. 10225522 (Tsinghua University), and the
Department of Energy under Contract No. DE-FG02-04ER41291 (U.
Hawaii).

\end{document}